\documentclass[final,3p]{elsarticle}

\usepackage{amsmath,amsfonts}
\usepackage{algorithmic}
\usepackage{algorithm}
\usepackage{array}
\usepackage[caption=false,font=normalsize,labelfont=rm,textfont=rm]{subfig}
\usepackage{textcomp}
\usepackage{stfloats}
\usepackage{url}
\usepackage{verbatim}
\usepackage{graphicx}
\usepackage{booktabs,ragged2e}
\usepackage{multirow}
\usepackage{multicol}
\usepackage{array}
\usepackage{colortbl}
\usepackage{tabularx}
\usepackage[flushleft]{threeparttable}
\usepackage{makecell}

\begin{document}


\title{A General Framework for Load Forecasting based on Pre-trained Large Language Model}

\author[label1]{Mingyang Gao} 
\author[label1]{Suyang Zhou} 
\author[label1]{Wei Gu} 
\author[label1]{Zhi Wu} 
\author[label1]{Haiquan Liu} 
\author[label2]{Aihua Zhou} 

\affiliation[label1]{organization={Southeast University},
            city={Nanjing},
            state={Jiangsu},
            country={China},
}

\affiliation[label2]{organization={China Electric Power Research Institute},
            city={Nanjing},
            state={Jiangsu},
            country={China}
}

\begin{abstract}
  Accurate load forecasting is crucial for maintaining the power balance between generators and consumers, particularly with the increasing integration of renewable energy sources, which introduce significant intermittent volatility. 
  With the advancement of data-driven methods, machine learning and deep learning models have become the predominant approaches for load forecasting tasks. 
  In recent years, pre-trained large language models (LLMs) have achieved significant progress, demonstrating superior performance across various fields. 
  This paper proposes a load forecasting method based on LLMs, offering not only precise predictive capabilities but also broad and flexible applicability. 
  Additionally, a data modeling method is introduced to effectively transform load sequence data into natural language suitable for LLM training. 
  Furthermore, a data enhancement strategy is designed to mitigate the impact of LLM hallucinations on forecasting results. 
  The effectiveness of the proposed method is validated using two real-world datasets. 
  Compared to existing methods, our approach demonstrates state-of-the-art performance across all validation metrics.
\end{abstract}



\maketitle

\section{Introduction}
\label{intro}
\subsection{Motivation}
Load forecasting has been playing an important role in maintaining the stability of the modern power system \cite{kundur2007power,7105972}. 
With accurate forecasting, the power system can optimize the integration of variable renewable energy sources \cite{brouwer2014impacts, guerra2022facing}.
As the advancement of data-driven algorithms, machine learning and deep learning-based methods have become the predominant approaches for load forecasting tasks, owing to their exceptional performance \cite{cai2019day}.

More recently, large language models (LLMs) have emerged, demonstrating strong accuracy and flexibility across various research in natural language process (NLP) tasks \cite{wang2022pre,min2023recent}. 
The attention mechanism in LLMs models have been proven to be effective in capturing the long-range dependencies in time series data, which is beneficial for load forecasting tasks \cite{fazlipour2022deep}.
However, there is currently no research utilizing LLMs for load forecasting tasks. 
The challenge lies in effectively considering the characteristics of electrical loads in conjunction with the superior knowledge comprehension and reasoning capabilities of LLMs to enhance the accuracy and reliability of predictions. 
This remains a promising issue that requires further investigation.

\subsection{Literature Review}
Load forecasting is a task within the domain of time series forecasting, and various networks have been continuously adapted to promote the prediction accuracy, such as eXtreme Gradient Boosting (XGBoost) \cite{chen2016xgboost}, Long Short-Term Memory (LSTM) \cite{hochreiter1997long} and Transformers \cite{vaswani2017attention}.
Notably, the Transformer architecture and its subsequent variants have demonstrated superior performance.
Informer \cite{zhou2021informer} proposes the ProbSparse self-attention mechanism, which reduces time complexity while maintaining performance. 
It employs self-attention distilling to manage long input sequences effectively.
Autoformer \cite{wu2021autoformer} introduces an auto-correlation mechanism by utilizing the pre-processing convention of series decomposition, transforming it into a fundamental component within models.
Fedformer \cite{zhou2022fedformer} introduces an approach that integrates Transformer models with seasonal-trend decomposition techniques and leverage the typically sparse representation of time series data in the Fourier transform.
Additionally, in DLinear \cite{zeng2023transformers}, researchers argue that accuracy is not primarily determined by the network architecture. 
Instead, the decomposition and processing of data significantly enhance the accuracy of the predictive models. 
By employing a simple fully-connected network with decomposition techniques , they are able to achieve similarly satisfactory prediction accuracy.

Consequently, beyond the model itself, the quality and distribution of data used for training above models also make a difference to the forecasting accuracy. 
To ensure the model performs optimally on general task, data modeling and feature engineering strategies designed for specific model are often proposed at the same time \cite{liu2019data}.
Reference \cite{abbasi2019short} adapts a XGBoost-based scheme for electricity load forecasting through increasing number of features available and converting daily electricity load information into weekly load information. 
In \cite{kong2017short, kong2017behaviour}, a residential load forecasting framework based on the LSTM is described with an customer-wise level data analysis.

Some ongoing research are already applying LLMs on time series forecasting tasks and obtain competitive forecasting results.
Reference \cite{xue2023promptcast} introduces a prompt-based learning paradigm for time series forecasting based on LLMs and shows superior performance across three distinct scenarios.
However, data missing is observed during the prediction process. This issue arises from the hallucination problem inherent in LLMs \cite{xu2024hallucination,ye2023cognitive}.
In load forecasting tasks, the hallucination may lead to extremely inaccurate predictions or missing values in the output sequence, but there is few research on how to effectively solve the problem. 
Reference \cite{jin2023time} also employs LLMs as the predictor, but they keeps the parameter of LLMs static and completes the training by updating the forward reprogramming layer.
Reference \cite{WU2024124034} leverages the reasoning ability of LLMs for accurate wind speed forecasting with spatio-temporal information.

\subsection{Contributions and Paper Organization}
In order to resolve the above deficiencies, this paper proposes a load forecasting framework based on pre-trained LLMs, leveraging its flexibility and generalizability to achieve more accurate results on multi-time-scale and multi-scenario datasets. 
Also, the paper introduces a dataset modeling method that enables LLMs to perform effectively. 

The specific contributions of this research are as follows:
\begin{enumerate}
  \item{We propose a general and flexible load forecasting method based on pre-trained LLMs. The proposed method can be applied to multi-timescale and multi-scenario load forecasting tasks.}
  \item{A dataset formulation method that combine language with statistical information is introduced to better leverage the predictive capabilities of LLMs.}
  \item{A data enhancement method is devised for solving the hallucination problems of LLMs by separating numerical sequence with language descriptions.}
  \item{The effectiveness of the proposed method is validated across open-sources and real-world load forecasting datasets with different time scales. Compared with existing load forecasting methods, the superiority and adaptability of the proposed framework is clearly proved.}
\end{enumerate}

The paper is structured as follows:
Section \ref{sec:dataset-df} outlines the dataset modeling approach for load forecasting tasks utilizing pre-trained LLMs. 
Section \ref{sec:model-arch} details the proposed load forecasting framework, encompassing the backbone models employed, training strategies, and evaluation metrics. 
Section \ref{sec:case-study} presents case studies to validate the effectiveness of the proposed methods. 
Finally, Section \ref{sec:conclusion} concludes the paper and discusses potential directions for future research.

\begin{table*}[ht]
  \centering
  \scriptsize
  \begin{threeparttable}
  \caption{The example of dataset based on proposed method}
  \label{tab:dataset_exps}
  \begin{tabular}{c c c c}
    \toprule
    Input Data & ${L\times d}$ & Example & Ground-truth\\
  \cmidrule{1-4}
    \multirow{1}{*}{$X_{text}$} & \makecell[l]{ELFD: 7$\times$1 \\ ICLD: 24$\times$1} & \makecell[l]{The electricity consumption of each day is as follows,\\ 29979,29415,27958,25579,28112,29664,29516kWh.\\ What is the daily consumption of next week?} &  \makecell[l]{The electricity consumption of each day\\ is as follows, 22992,21895,26303,28286,\\28727,26488,24839kWh.} \\
  \cmidrule{1-4}
    \multirow{1}{*}{$X_{ts}$} & \makecell[l]{ELFD: 7$\times$1 \\ ICLD: 24$\times$1} & \makecell[l]{The electricity consumption of each day is as follows,\\ 29979,29415,27958,25579,28112,29664, 29516kWh. \\ The maximum value is 32123, the minimun value is\\ 20321, the average value is 28603. What is the daily\\ consumption of next week?} & \makecell[l]{The electricity consumption of each day\\ is as follows,22992,21895,26303,28286,\\28727,26488,24839kWh.} \\
  \cmidrule{1-4}
    \multirow{1}{*}{$X_{ets}$} & \makecell[l]{ELFD: 7$\times$1 \\ ICLD: 24$\times$1} & \makecell[l]{The electricity consumption of day one is 29979,\\ the electricity consumption of day two is 29415,\\ the electricity consumption of day three is 27958,\\ the electricity consumption of day four is 25579,\\ the electricity consumption of day five is 28112,\\ the electricity consumption of day six is 29664,\\ the electricity consumption of day seven is 29516.\\ The maximum value is 32123, the minimun value\\ is 20321, the average value is 28603. What is the\\ daily consumption of next week?} & \makecell[l]{The electricity consumption of day one is 22992, \\the electricity consumption of day two is 21895,\\ the electricity consumption of day three is 26303,\\ the electricity consumption of day four is 28286,\\ the electricity consumption of day five is 28727, \\ the electricity consumption of day six is 26488,\\ the electricity consumption of day seven is 24839.} \\
  \bottomrule
  \end{tabular}
  \begin{tablenotes}
    \footnotesize
    \item For datasets in other languages, we use Google Translate to generate corresponding input data and Ground-truth in the identical format.
  \end{tablenotes}
\end{threeparttable}
\end{table*}

\section{Dataset Formulation and Enhancement}
\label{sec:dataset-df}
In this section, we present a method for creating datasets for language models. 
Starting with converting numerical data into textual data, we will detail the approach through which we can effectively used the data in.
Moreover, a technique to address the hallucination phenomenon is also introduced. 
To emphasize, the proposed dataset modeling method is applicable to all load forecasting tasks based on language models.

\subsection{Combine Language with Statistical Information}
In common load forecasting tasks, historical load data are always employed as the input for forecasting.
The input data are typically modelled into a continuous sequence $X\in \mathbb{R}^{L\times  d}$, where $L$ and $d$ represents the length and dimension of the sequence, respectively.
Since data is required to be input in text format for language models, we propose a dataset modeling method that convert numerical sequence into natural language expression $X_{text}$ as described below:
\begin{equation}
  X_{text} = \mathbb{S}(X) =  \{ \mathbb{S}(x_{1}) \ ... \ \mathbb{S}(x_{i})\ ... \ \mathbb{S}(x_{n})\} \qquad for \ 1\leq i \leq n \hfill
\end{equation}
where $x_{i}$ is the $i$-th data in the input sequence,
$\mathbb{R}$ represents the set of real number,
$\mathbb{S}$ stands for the transformation from real number to text.

Additionally, to further exploit the advantages of textual expression and inspired by references \cite{zeng2023transformers}, which demonstrate that data decomposition and processing significantly enhance predictive model accuracy,
we introduce statistical information $X_{stat}$ to enhance the feature dimension of the input data, denoted as $X_{ts}$.

The statistical information includes maximum, minimum, and average values to cover the local and global . 
Specifically, we use the maximum and minimum values within the range of $N_{obs}$ steps before the predicted time to model global features,
and represent the local features with the average value of the input sequence.
\begin{equation}
    \left\{\begin{array}{l}
    X_{ts} = \{X_{text}, X_{stat} \} \hfill\hfill
    \\ 
    X_{stat} = \{ Max(X_{obs}),Min(X_{obs}),Average(X) \}\hfill\hfill
    \end{array}\right.
\end{equation}
where $X_{ts}$ represents the input with statistical information,
$X_{stat}$ is the statistical information with language descriptions including the maximum, minimum and average value of $X$,
$X_{obs}$ demonstrate the historical load data within the range of $N_{obs}$ time-steps before the predicted time.
\subsection{Separate Numerical Sequence with Language}
The causes of hallucination in load forecasting tasks, such as missing data or generating extra data, can be attributed to two primary aspects:
1) during the conversion of numerical data to textual descriptions, the lengths of loaded data stored in string format exhibit inconsistency.
2) the pre-training parameters of LLMs are derived from training on natural language, thereby lacking the capability to effectively recognize purely numerical values.

Taking advantages of LLM's sensitivity to language descriptions, this section proposes a data enhancement method that separates numerical data with textual information. 
The enhanced input dataset $X_{ets}$ is constructed based on $X^{*}_{text}$ as shown below:
\begin{equation}
  \left\{\begin{array}{l}
  X_{ets} = \{X^{*}_{text}, X_{stat} \} \hfill\hfill
  \\
  X^{*}_{text} = \{ (t_{1}, x_{1}) \ ... \ (t_{i}, x_{i}) \ ... \ (t_{n}, x_{n}) \} \qquad for \ 1\leq i \leq n  \hfill
  \\ 
  X_{stat} = \{ Max(X_{obs}),Min(X_{obs}), Average(X)\}\hfill\hfill
  \end{array}\right.
\end{equation}
where $X^{*}_{text}$ is the textual expression of the numerical sequence with time information,
$t_{i}$ is the $i$-th corresponding time-steps in textual expression of $x_{i}$.

Given that the output data from LLMs also exists in text form, 
the textual ground-truth is necessary for the training process consequently.
Following the same process for each input format of $X$, we generate the corresponding ground-truth $Y_{gt}$.

\subsection{Dataset for Forecasting}
To evaluate the generality and accuracy of the proposed methods in load forecasting tasks, 
we selected the following two real-world datasets at different time scales to perform our research.

\textbf{Industrial Clients Load Dataset (ICLD)}:This real-world dataset comprises around 9000 daily load data on the electricity consumption of the 10 industrial clients from June 1st, 2018 to June 25th, 2021.
All data is collected from a real-world city-level power system in east China. The time length of the training/validation/test set is 24/6/6 months, respectively.
The average and standard deviation value of ICLD is 3695.10 and 2334.11.

\textbf{Electricity Load Forecasting Dataset (ELFD)}:This is an open-source dataset available on Kaggle\footnote{kaggle.com/datasets/saurabhshahane/electricity-load-forecasting/data}, covering over 40,000 hourly load data for the Panama region from January 31st, 2015 to June 10th, 2020.
The time length of the training/validation/test set is 48/12/6 months, respectively.
The average and standard deviation value of ICLD is 1184.82 and 192.26.

\begin{figure*}[t]%
  \centering
  \footnotesize
  \subfloat[ICLD]{
      \label{Dataset-fig1}
      \includegraphics[width=8cm,height=6cm]{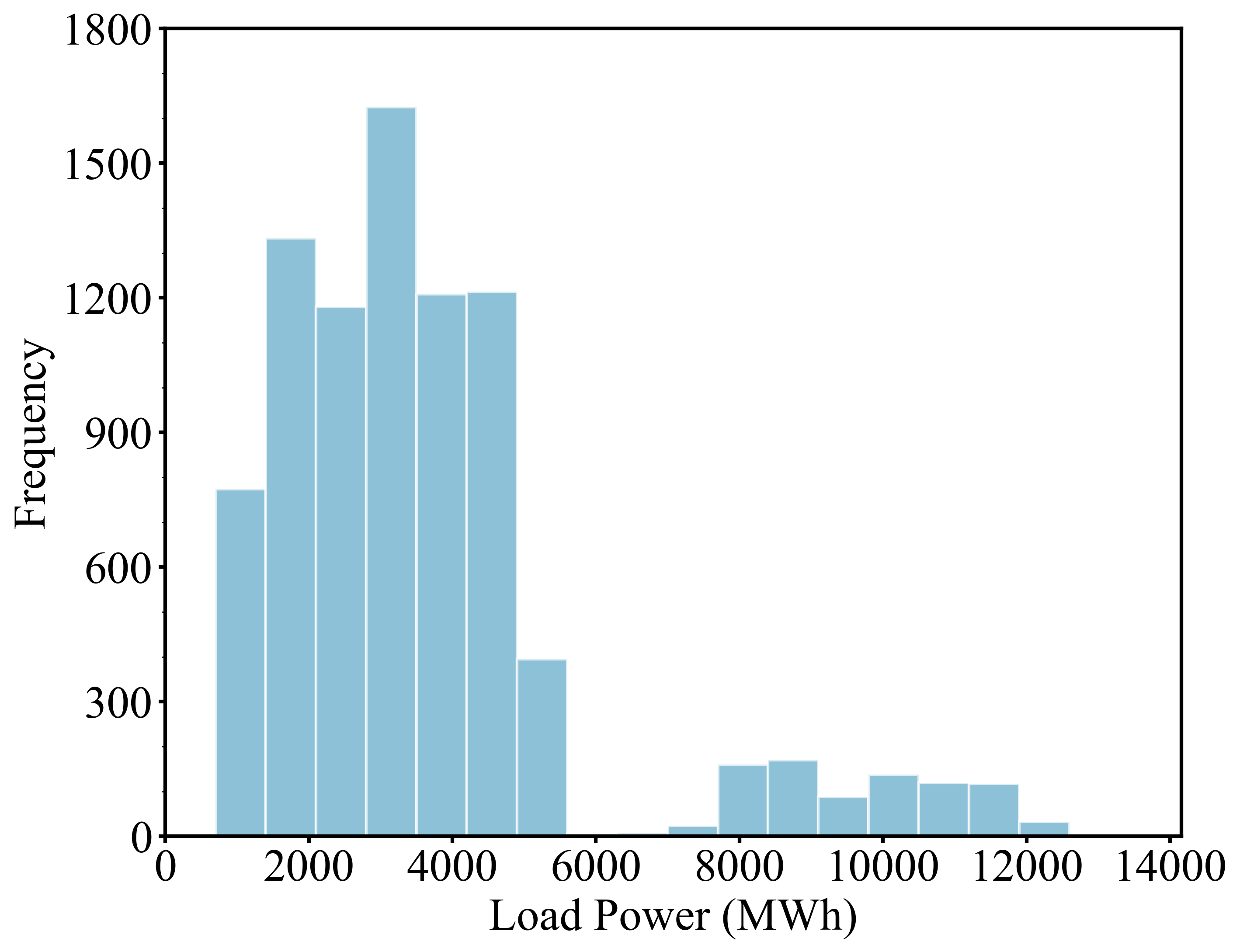}
      }
  \subfloat[ELFD]{
      \label{Dataset-fig2}
      \includegraphics[width=8cm,height=6cm]{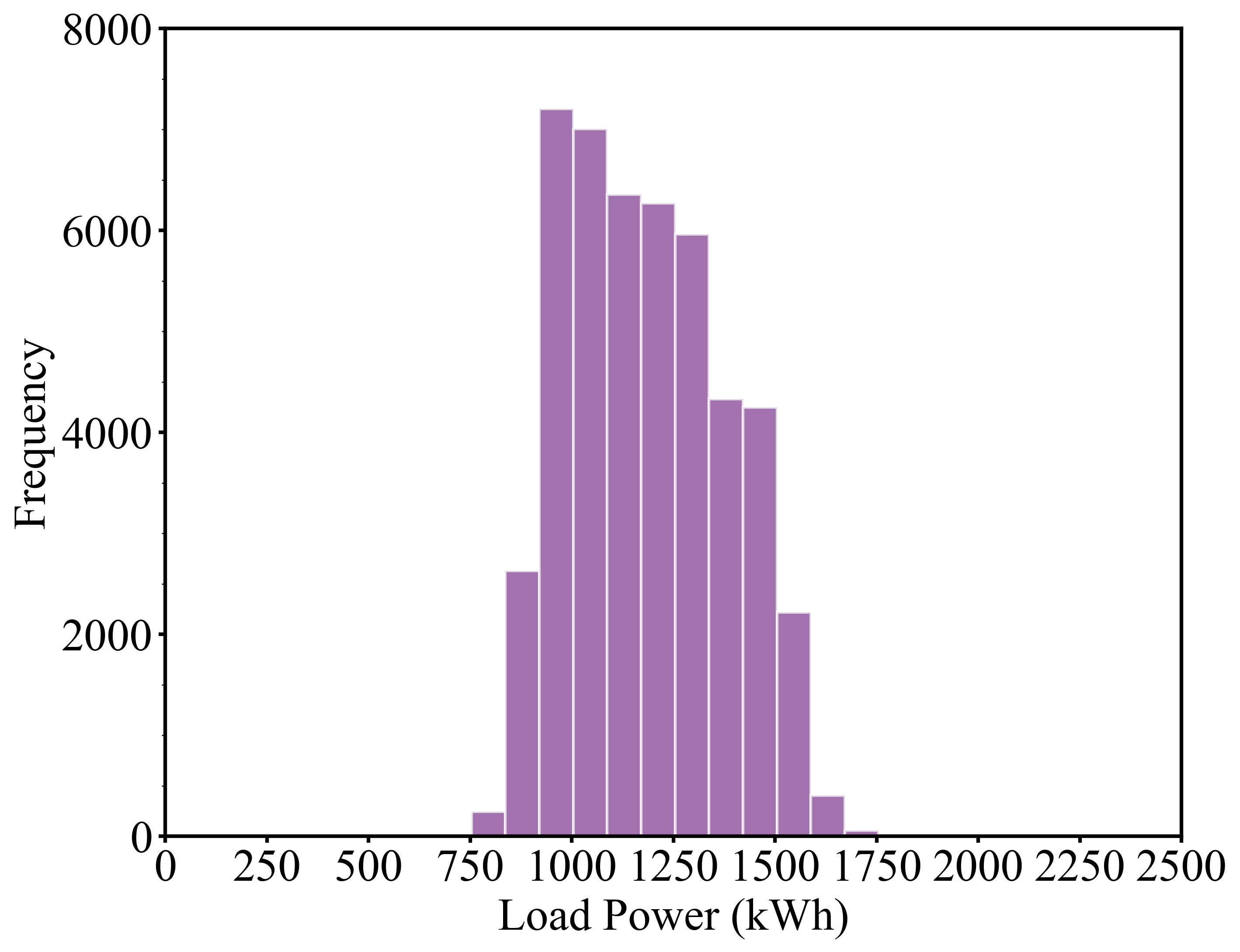}
      }
  \captionsetup{name={Fig.}, labelsep=space, justification=raggedright, singlelinecheck=false} 
  \caption{Distribution of involved load forecasting datasets}
  \label{fig:Distribution-Datasets}
  \vspace{-10pt}
\end{figure*}

The distribution of two dataset is visualized in Figure \ref{fig:Distribution-Datasets}. 
Dataset with detailed examples under the strategies established in this section is shown in Table \ref{tab:dataset_exps}.
The effectiveness of proposed methods above are validated in Section \ref{sec:case-study}.

\section{Proposed Framework}

\label{sec:model-arch}
In this section, we provide a detailed introduction to the basic structure of our proposed prediction framework and the LLMs used to complete the prediction task. 
Additionally, we detail training methods for different LLMs and list the metrics used to evaluate their forecasting results.
\subsection{Multi-Head Attention Mechanism within LLMs}
The Multi-Head Attention mechanism comprises three primary components: linear projections, scaled dot-product attention, and concatenation followed by linear transformation, 
which enables the model to process input data from multiple perspectives simultaneously, enhancing its ability to capture complex patterns and relationships within the sequence.

\textbf{1)Linear Projections:} For each attention head, the input vectors are linearly projected into three distinct spaces to create queries $Q_{i}$, keys $K_{i}$, and values $V_{i}$, allowing each head to focus on different aspects of the input data.
\begin{equation}
    Q_{i} = QW_{i}^{Q} \qquad K_{i} = KW_{i}^{K} \qquad V_{i} = VW_{i}^{V}
\end{equation}
where $W_{i}^{Q}$, $W_{i}^{K}$, and $W_{i}^{V}$ are the linear projection matrices for each head.

\textbf{2)Scaled Dot-Product Attention:} Each attention head computes attention scores and passed through a softmax function to obtain attention weights, which are used to compute a weighted sum of the value vectors.
\begin{equation}
    head_{i} = \text{Attention}(Q_{i},K_{i},V_{i}) = \text{softmax}(\frac{Q_{i}K_{i}^{T}}{\sqrt[]{d_{k}}})V_{i}
\end{equation}
where $head_{i}$ denotes the output of the $i$-th attention head, $d_{k}$ is the dimension of the key vectors and $1 / \sqrt[]{d_{k}}$ stands for the scaling factor.

\textbf{3)Concatenation and Linear Transformation:} The outputs from all attention heads are concatenated and passed through a final linear transformation, which combines the diverse information captured by each head into a single output representation.
\begin{equation}
    \text{MultiHead}(Q_{i},K_{i},V_{i}) = \text{Concat}(head_{1},...,head_{h})W^{O}
\end{equation}
where $W^{O}$ is the linear transformation matrix for the output of network, $h$ is the number of attention heads.

\subsection{LLMs for Load Forecasting}
LLMs could be structurally categorized into three types:

\textbf{Encoder-Only Models:} Represented by BERT \cite{DBLP:journals/corr/abs-1810-04805}, these models learn bidirectional context encoders through masked language modeling. 
The training objective involves randomly masking parts of the text and predicting the masked words. 
This architecture is mainly suitable for tasks that do not require sequence generation but instead need to encode and process input, such as text classification and sentiment analysis.

\textbf{Decoder-Only Models:} Represented by GPT \cite{Radford2019LanguageMA} and BLOOM \cite{Scao2022BLOOMA1}, these models are typically used for sequence generation tasks and known as generative model. 
It generates sequences directly from the input and perform unsupervised pre-training. 
However, they require tremendous training data to improve the quality and diversity of generated text.

\textbf{Encoder-Decoder Models:} Represented by T5 \cite{zhang2021mengzi} and BART \cite{DBLP:journals/corr/abs-1910-13461}, these models use an encoder to process the input sequence, extracting features and semantic information, and a decoder to generate the corresponding output sequence. 
Known as sequence-to-sequence model, it experts in handling the relationship between input and output sequences, improving accuracy in tasks like machine translation and dialogue generation. 

\begin{figure*}[ht]
  \centering
  \includegraphics[width=\textwidth]{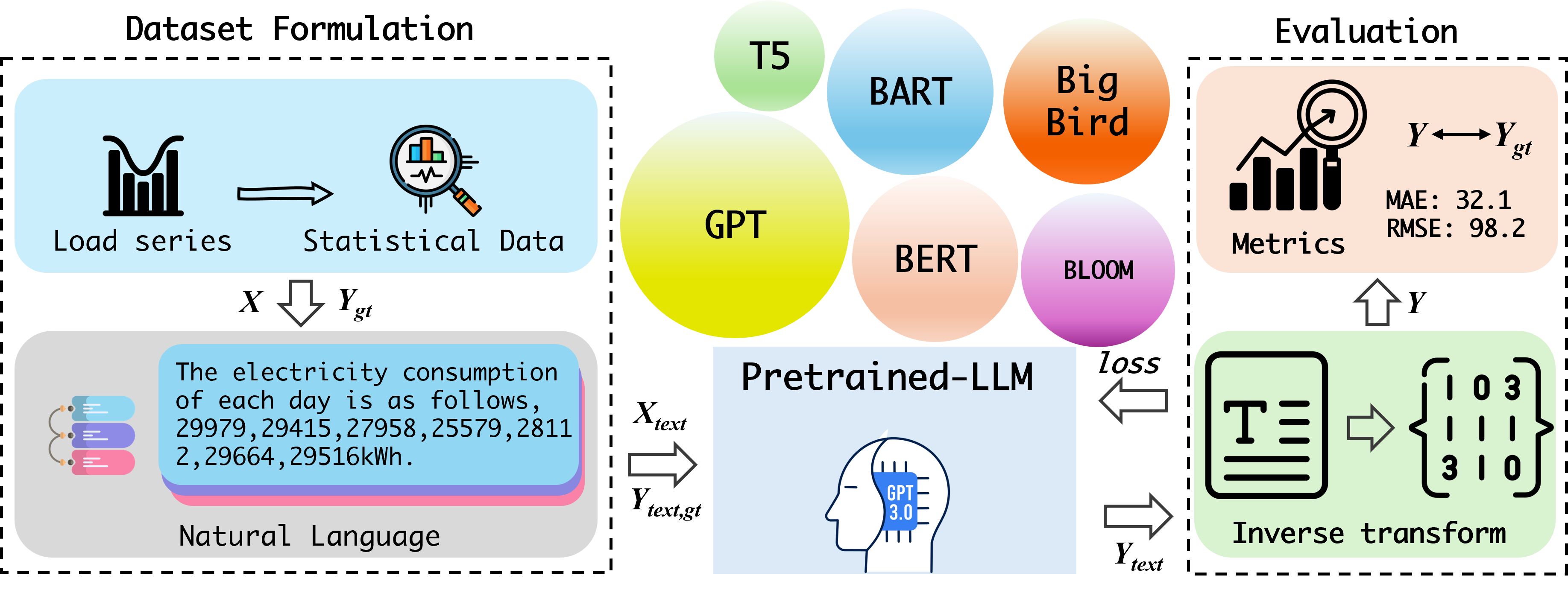}
  \captionsetup{name={Fig.}, labelsep=space, justification=raggedright, singlelinecheck=false} 
  \caption{The architecture of the proposed framework.}
 \label{main-structure}
 \vspace{-10pt}
\end{figure*}

\begin{table}[ht]
	\centering
  \scriptsize
  \caption{The LLM Models used in the Proposed Framework}
  \label{tab:LLMs}
  \begin{tabular}{l c c c}  
  \toprule
    Model & \makecell[c]{Pre-trained Language} & Access Key & Model Size\\
  \cmidrule{1-4}
    \makecell[l]{GPT2} & \makecell[c]{English} & \makecell[c]{openai-community/gpt2} & \makecell[c]{548MB}\\
    \addlinespace
    \makecell[l]{BART \\\addlinespace BART-CN} & \makecell[c]{English \\\addlinespace Chinese} & \makecell[c]{facebook/bart-base \\\addlinespace fnlp/bart-base-chinese} & \makecell[c]{558MB \\ \addlinespace561MB}\\
    \addlinespace
    \makecell[l]{T5 \\\addlinespace Mengzi-T5} & \makecell[c]{English \\\addlinespace Chinese} & \makecell[c]{google-t5/t5-base \\\addlinespace Langboat/mengzi-t5-base} & \makecell[c]{892MB \\\addlinespace 990MB}\\
    \addlinespace
    \makecell[l]{BigBird} & \makecell[c]{English} & \makecell[c]{google/bigbird-pegasus-\\large-arxiv} & \makecell[c]{2.3GB}\\
    \addlinespace
    \makecell[l]{BLOOM \\\addlinespace BLOOM-CN} & \makecell[c]{English \\\addlinespace Chinese} & \makecell[c]{bigscience/bloom-1b7 \\\addlinespace Langboat/bloom-1b4-zh} & \makecell[c]{3.4GB \\\addlinespace 5.6GB}\\
  \bottomrule
\end{tabular}
\end{table}

Depending on the characteristics of load forecasting tasks, 
we primarily considers LLMs based on Decoder-only and Encoder-Decoder architectures.
In light of the models discussed in references \cite{xue2023promptcast, WU2024124034}, we selected several open-source LLMs for load forecasting. The configurations for these models are presented in Table \ref{tab:LLMs}. 
Furthermore, LLMs trained in different languages are selected to verify whether the forecasting result is influenced by natural language expression.
\subsection{Training Strategies for LLMs}
\subsubsection{Parameter-Efficient Fine-Tuning (PEFT)}
Large language models pre-trained for general tasks, encode a comprehensive understanding of knowledge within their pre-trained parameters. 
Consequently, training these models completely on specialized datasets will destroy the distribution pattern of pre-trained parameters, reducing their feasibility in text comprehension. 
Therefore, we adoption PEFT method with the Low-Rank Adaptation of Large Language Models (LoRA) technique to delicately fine-tune model parameters \cite{ding2023parameter}.
In this method, we use low-rank decomposition to simulate parameter changes based on the original model's parameter distribution, thereby indirectly training a large model with a minimal number of parameters.
We process the selected parameter matrix $W_{d \times k}$ from the original model as follows:
\begin{eqnarray}
  W_{d \times k} = U_{d \times r} \cdot V_{r \times k} \qquad r \ll d,k
\end{eqnarray}
where $r$ is the Low-rank coefficient, $U$ and $V$ are the low-rank matrices.

In our research, the parameters selected for PEFT are linear transform layers and attention layers. The total amount of trainable parameters takes up to 10\% of the original model.

\subsubsection{Full Parameter Training}
Fully parameterized training method under our proposed framework is also served to train LLMs with the proposed dataset. 
While this approach trades off original problem-solving capabilities for the utilization of pre-trained parameters, it demonstrates notable efficacy in load forecasting tasks.

\subsection{Evaluation Method and Metrics}
For the model's prediction results, we mostly care about the accuracy of the numerical sequence within natural language. 
According to the format setting of ground-truth in Section \ref{sec:dataset-df}, we can easily extract the data sequence from the text, 
with which we can calculate the forecasting accuracy to analyse the performance of the model.

Hallucination Rate is proposed to evaluate the hallucination in the forecasting results.
Together with the Mean Absolute Error (MAE) and Root Mean Square Error (RMSE), three metrics are served as evaluation metrics in our research and defined as follows, 

\begin{equation}
  \left\{\begin{array}{l}
    H = n_{h} / {N}
    \\
    MAE = \frac{1}{N}  \sum_{i=1}^{N}\left | Y_{i} - Y_{i,gt} \right | 
    \\
    RMSE =  \sqrt[]{\frac{1}{N}  \sum_{i=1}^{N}\left ( Y_{i} - Y_{i,gt} \right )^2} 

  \end{array}\right.
\end{equation}
where $N$ is the number of samples, $Y_{i}$ is the $i$-th predicted result, $Y_{i,gt}$ is the corresponding ground-truth, $H \in (0 , 1)$ is the Hallucination Rate and $n_{h}$ is the number of hallucination samples.

The framework of our work is depicted in Figure \ref{main-structure}.
To fully leverage the pre-trained parameters in large models, we adopt diverse training approaches for various LLMs, 
aiming to achieve optimal prediction results while maintaining training efficiency.
\begin{table}[ht]
  \centering
  \caption{Hyperparameters of Proposed and Comparison Methods}
  \label{tab:hyper}
  \begin{tabular}{ c l l}
  \toprule
    Method & Hyperparameters & Value \\
  \hline
  \multirow{1}{*}{LLM} & \makecell[l]{batch size \\ Learning rate \\ Input length of ICLD \\ Output length of ICLD \\ Input length of ELFD \\ Output length of ELFD \\LoRA coefficient \\ LoRA alpha\\ LoRA dropout} & \makecell[l]{32 \\ $5e^{-5}$ \\ 7 \\ 7 \\ 24 \\ 24 \\8\\ 32\\ 0.1}\\
  \hline
  \multirow{1}{*}{XGBoost} & \makecell[l]{ Number of estimators\\ Learning rate\\ Max depth} &  \makecell[l]{160 \\ 0.001 \\ 10}\\
  \hline
  \multirow{1}{*}{LSTM} & \makecell[l]{Number of layers\\ Hidden size\\ Dropout rate \\ Batch size \\Learning rate} & \makecell[l]{10 \\ 128 \\ 0.2 \\ 32 \\ 0.001} \\
  \hline
  \multirow{1}{*}{X-former} & \makecell[l]{ Number of heads\\ Moving average step\\ Number of Enc/Dec layers \\ Batch size \\Learning rate} & \makecell[l]{8 \\ 12 \\ 2/1 \\ 32 \\ 0.001} \\
  \hline
  \multirow{1}{*}{Dlinear} & \makecell[l]{Kernel size \\ Individual \\ Batch size \\Learning rate} & \makecell[l]{ 25 \\ 0 \\ 32 \\ 0.001} \\
  \bottomrule
\end{tabular}
\end{table}

\begin{figure}[H]
  \centering
  \footnotesize
  \includegraphics[width=\textwidth]{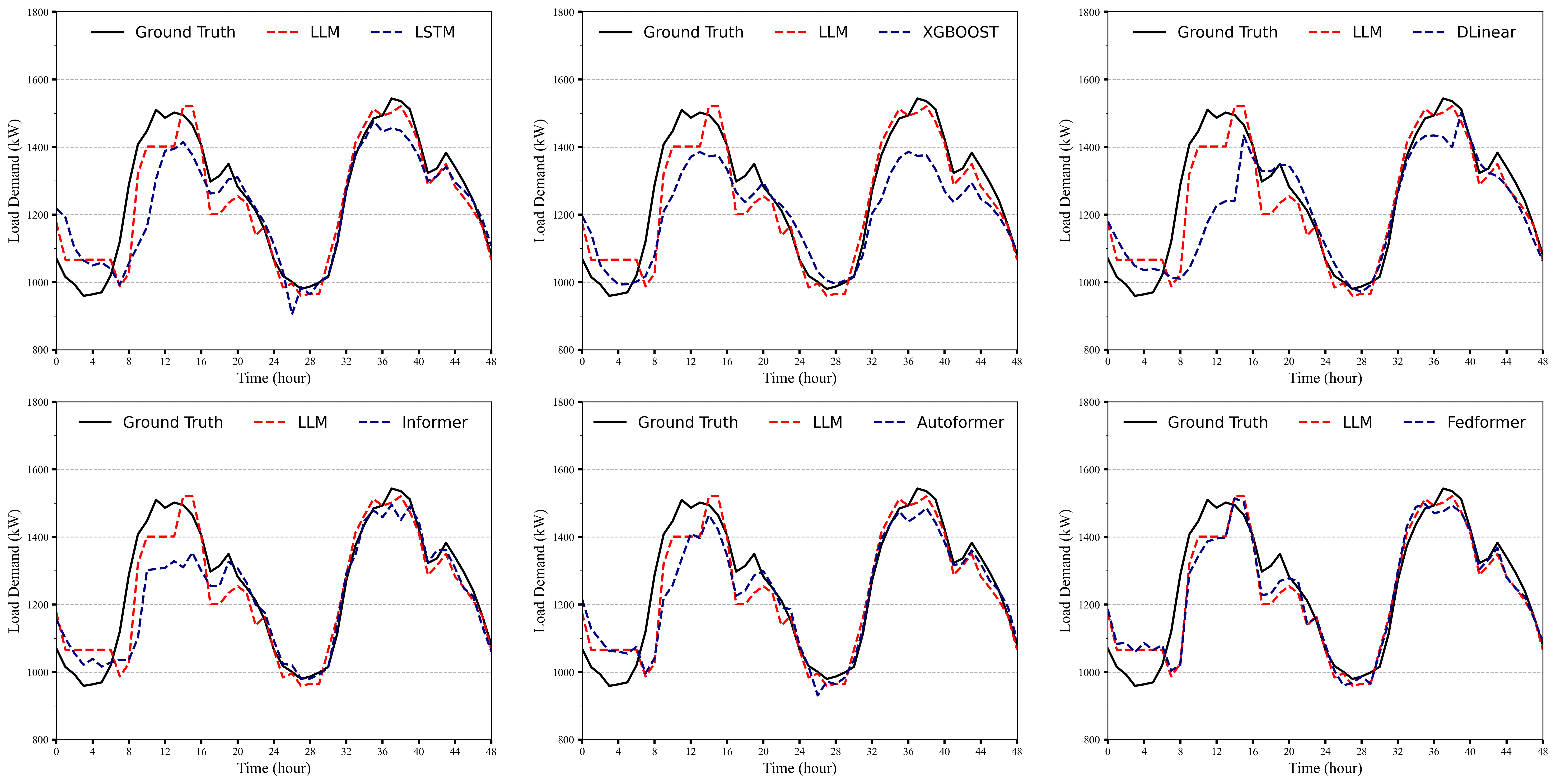}
  \captionsetup{name={Fig.}, labelsep=space, justification=raggedright, singlelinecheck=false} 
  \caption{Forecasting results comparison of the proposed method with other models on ELFD.}
 \label{fig:kg_24hours}
 \vspace{-10pt}
\end{figure}

\section{Case study}
\label{sec:case-study}
In this section, we will validate the effectiveness of the proposed methods. 
Firstly, we will demonstrate the physical environment and hyperparameter configurations employed during training.
Secondly, we will apply the forecasting framework to two datasets introduced in Section \ref{sec:dataset-df}. 
Mengzi-T5 model as a representation of LLMs, will undergo an in-depth evaluation of its performance and statistical outcomes compared with traditional methods.
Furthermore, the various LLMs mentioned in Section \ref{sec:model-arch} will be tested to confirm their capabilities in the prediction task.

\subsection{Parameters Configuration}
Our model is implemented using PyTorch and Transformers from HuggingFace, with all experiments conducted on NVIDIA 4090-24G GPUs.
All of our models can be accessed with the Access Key in Table \ref{tab:LLMs} from HuggingFace Model Hub \cite{DBLP:journals/corr/abs-1910-03771}.
The hyperparameters of the proposed framework and comparison methods are shown in Table \ref{tab:hyper}.

\subsection{Case1: Forecasting Results of Different Time-scale Datasets}
In the validation sets of ICLD and ELFD, with lengths of 6 and 12 months respectively, we calculate the Hallucination Rate, MAE and RMSE of the predicted data. 
The hallucination of LLMs may result in missing or excess issues with the results. 
To ensure the calculation of metrics, we address this problem as follows: 1)missing data is handled by supplementing it with zeros, 2)additional data is removed to keep the same length of all output sequences.
We employ GPT2 model for our framework and compare it with traditional methods including XGBoost, LSTM, Informer, Autoformer, Fedformer and DLinear.

As depicted in Table \ref{tab:cmp_res}, our method shows state-of-the-art performance compared to all traditional prediction methods. 
The LLM, LLM-ts, and LLM-ets methods are only different in the input data format corresponding to $X_{text}$, $X_{ts}$, and $X_{ets}$.
The prediction results suggest that integrating long-vision statistical information into the data enhances prediction accuracy. 
To provide a intuitional presentation, forecasting curves of the our framework compared with other methods are visualized in Figure \ref{fig:kg_24hours}.

Notably, the red data in Table \ref{tab:cmp_res} highlights the impact of the hallucination problem within forecasts, particularly in the ICLD dataset, where it results in a significantly higher RMSE than normal values. 
We confirm that this issue arises from a missing value in the predicted data sequence.
With $X_{ets}$ as the input data, the hallucination rate is reduced to zero, and the MAE and RMSE are also significantly improved.

The forecasting result demonstrates that without preprocessing the original data, the predictive capability of LLM is under-explored.
With the proposed method in Section \ref{sec:dataset-df}, LLMs can effectively eliminate hallucination and improve the forecasting accuracy.
In contrast to traditional model data preprocessing methods, the approach based on pre-trained LLMs leverages decomposition in a more straightforward and efficient way by simply incorporating statistical information into the prompts.
\begin{table}[H]
  \centering
  \small
  \caption{Comparison between Different Methods and Proposed Framework}
  \label{tab:cmp_res}
  \begin{tabular}{lc ccc|ccc}
    \toprule
    \multirow{2}{*}{Method} & \multirow{2}{*}{Input Data} & \multicolumn{3}{c|}{ICLD} &  \multicolumn{3}{c}{ELFD}\\ 
    \cmidrule{3-8}
      & & Hallucination Rate & MAE & RMSE & Hallucination Rate & MAE & RMSE \\
      \midrule
       XGBoost & $X$ &0 & 130.50 & 219.54 &0 & 83.35 & 113.68\\
       \belowrulesep=0pt \aboverulesep=0pt
       LSTM & $X$ & 0 & 172.44  & 294.69 & 0& 85.08 & 117.52 \\
       Informer & $X$ & 0 & 92.71  & 167.68 & 0& 63.15 & 78.51 \\
       Autoformer & $X$ & 0 & 87.89  & 150.18 & 0& 61.15 & 79.89 \\
       Fedformer & $X$ & 0 & 82.01  & 126.31 & 0& 57.15 & 74.37 \\
       Dlinear & $X$ & 0 & 99.40  & 189.73 & 0& 74.25 & 106.51 \\
       LLM & $X_{text}$ & \textcolor{red}{0.035} & 264.53  & \textcolor{red}{2987.19}  & \textcolor{red}{0.085} & 203.25 & \textcolor{red}{457.62} \\
       LLM-ts & $X_{ts}$ & \textcolor{red}{0.022} & 239.60  & \textcolor{red}{2182.43}  & \textcolor{red}{0.016} & 116.05 & \textcolor{red}{364.31} \\
       LLM-ets & $X_{ets}$ & \textbf{0}& \textbf{80.62} & \textbf{112.59}  & \textbf{0}& \textbf{52.44} & \textbf{73.03}\\
  \bottomrule
  \end{tabular}
\end{table}

\subsection{Case2: Forecasting Results based on Different LLMs Models}
We validated the generality of our approach on LLMs with different backbones given in Table \ref{tab:LLMs}.
As shown in Table \ref{tab:LLMs_cmp_res}, the proposed prediction framework consistently achieves low MAE and RMSE across different LLMs, and the GPT2 model shows the best prediction performance on both datasets.

Additionally, we conducted comparative experiments using two state-of-the-art LLM framework, GPT-4 and Claude 3.5. 
For these models, we obtained predictions by inputting prompts without any additional training. 
The results indicate that LLMs demonstrate a certain level of predictive capability even without being trained on specific datasets. 
However, even the most advanced commercially available LLMs, when not fine-tuned on task-specific datasets, underperform compared to models with fewer parameters. 
This result further validates the efficacy of our proposed methodology.


\begin{table}[H]
  \belowrulesep=0pt \aboverulesep=0pt
  \centering
  \caption{Comparison between Different LLM backbones}
  \label{tab:LLMs_cmp_res}
  \begin{tabular}{ l cc|cc}
    \toprule
    \multirow{2}{*}{Model} & \multicolumn{2}{c|}{ICLD} &  \multicolumn{2}{c}{ELFD}\\ 
    \cmidrule{2-5}
      & MAE & RMSE & MAE & RMSE \\
      \midrule
       GPT4o (Without training) &  151.58 & 178.05  & 86.27 & 95.76 \\
       Claude3.5 (Without training) &  103.80 & 151.92  & 82.61 & 100.14 \\
       GPT2  & 80.62 & 112.59 & \textbf{52.44} & \textbf{73.03}\\
       T5  & 84.59 & 139.05 & 59.65 & 80.23\\
       Mengzi-T5  & \textbf{80.10} & \textbf{104.01} & 60.01 & 83.87\\
       BART &  91.58 & 216.3  & 66.27 & 92.19 \\
       BART-CN & 93.37  & 159.02 & 67.31 & 90.51 \\
       BigBird & 102.93 & 187.23  & 65.16 & 88.47 \\
       BLOOM & 99.17  & 238.01  & 62.25 & 95.98 \\
       BLOOM-CN &  109.74  & 247.08 & 62.31 & 98.05 \\
  \bottomrule
  \end{tabular}
  \end{table}

\section{Conclusion}
\label{sec:conclusion}
In this paper, a general and flexible load forecasting framework based on pre-trained language models is proposed.
The following conclusions can be drawn:
\begin{enumerate}
\item A dataset formulation approach is established to convert sequence-formatted data into natural language to facilitate LLM training and language descriptions of statistical information is integrated for broaden the input feature dimension. 
\item A data enhancement method is accordingly proposed to address the hallucination problem of LLMs in load prediction tasks.
With the proper separation of numerical sequence and language descriptions, the hallucination rate is significantly reduced to 0\%.
\item The comprehensive predictive performance of our method is validated on two real-world datasets.
The MAE is reduced to 80.10 and 52.44 on ICLD and ELFD respectively, demonstrating superior prediction accuracy over existing methods.
\end{enumerate}

In future work, we aim to apply larger language models on load prediction problems.
We will focus on establishing datasets and developing training methods suitable for large language models, ensuring reliable load prediction while maximizing the utilization of pre-trained parameters.

\section*{CRediT Authorship Contribution Statement}
\textbf{Mingyang Gao:} Conceptualization, Methodology, Software, Formal analysis, Writing - Original draft preparation.
\textbf{Suyang Zhou:} Conceptualization, Investigation, Supervision, Writing - Reviewing and Editing, Funding acquisition.
\textbf{Wei Gu:} Resources, Validation, Data curation, Project administration.
\textbf{Zhi Wu:} Supervision, Writing - Reviewing and Editing.
\textbf{Haiquan Liu:} Supervision.
\textbf{Aihua Zhou} Supervision.

\section*{Data availability}
Public datasets are used. The data can be accessed with the following URL: \url{https://www.kaggle.com/datasets/saurabhshahane/electricity-load-forecasting/data}.

\section*{Declaration of Competing Interest}
The authors declare that they have no known competing financial interests or personal relationships that could have appeared to influence the work reported in this paper.

\section*{Acknowledgment}
This work is supported by the Science and Technology Project of State Grid under the grant number 5700-202458232A-1-1-ZN \textit{(Corresponding author: Suyang Zhou).}

\bibliographystyle{elsarticle-num}
\bibliography{References}

\end{document}